# Bounding the Sum of Square Roots via Lattice Reduction *

Qi Cheng[1] and Xianmeng Meng[2] and Celi Sun[1] and Jiazhe Chen[2]

[1] School of Computer Science
The University of Oklahoma
Norman, OK 73019, USA.
Email: {`qcheng, scl`}@cs.ou.edu.
[2] Lab of Cryptographic Technology and Information Security
Shandong University
Jinan 250100, P.R. China.

**Abstract.** Let $k$ and $n$ be positive integers. Define $R(n,k)$ to be the minimum positive value of

$$|e_i\sqrt{s_1} + e_2\sqrt{s_2} + \cdots + e_k\sqrt{s_k} - t|$$

where $s_1, s_2, \cdots, s_k$ are positive integers no larger than $n$, $t$ is an integer and $e_i \in \{1, 0, -1\}$ for all $1 \le i \le k$. It is important in computational geometry to determine a good lower and upper bound of $R(n,k)$. In this paper we show that this problem is closely related to the shortest vector problem in certain integral lattices and present an algorithm to find lower bounds based on lattice reduction algorithms. Although we can only prove an exponential time upper bound for the algorithm, it is efficient for large $k$ when an exhaustive search for the minimum value is clearly infeasible. It produces lower bounds much better than the root separation technique does. Based on numerical data, we formulate a conjecture on the length of the shortest nonzero vector in the lattice, whose validation implies that our algorithm runs in polynomial time and the problem of comparing two sums of square roots of small integers can be solved in polynomial time. As a side result, we obtain constructive upper bounds for $R(n,k)$ when $n$ is much smaller than $2^{2k}$.

## 1 Introduction

Comparing sums of square roots of integers is a famous open problem in computational geometry and numerical analysis. It arises when we need to compare the length of two polygonal paths in a Euclidean space. The problem takes another form when one compares a sum of square roots with an integer. This problem is not known to be in NP. In fact, PSPACE is the smallest well studied complexity class that provably contains this problem [11]. In practice, however, it can usually be solved quickly.

* This research is partially supported by NSF Career Award CCR-0237845 of USA and by Project 973 (no: 2007CB807903 and no: 2007CB807902) of China.

**Definition 1.** *Define $r_1(n,k)$ to be the minimum positive value of*

$$\left|\sqrt{s_1} + \sqrt{s_2} + \cdots + \sqrt{s_{\lfloor k/2 \rfloor}} - \sqrt{s_{\lfloor k/2 \rfloor + 1}} - \cdots - \sqrt{s_k}\right|$$

*where $s_1, s_2, \cdots, s_k$ are positive integers less than or equal to $n$. Define $r_2(n,k)$ to be the minimum positive value of*

$$|\sqrt{s_1} + \sqrt{s_2} + \cdots + \sqrt{s_k} - t|$$

*where $s_1, s_2, \cdots, s_k$ are positive integers less than or equal to $n$ and $t$ is an integer. Define $R(n,k)$ to be the minimum positive value of*

$$|e_1\sqrt{s_1} + e_2\sqrt{s_2} + \cdots + e_k\sqrt{s_k} - t|$$

*where $s_1, s_2, \cdots, s_k$ are positive integers no larger than $n$, $t$ is an integer and $e_i \in \{1, 0, -1\}$ for all $1 \leq i \leq k$.*

For example, we have that

$$r_1(3,3) = \left|\sqrt{3} - \sqrt{1} - \sqrt{1}\right| \approx 0.268,$$

$$r_2(3,3) = \left|\sqrt{3} + \sqrt{2} + \sqrt{1} - 4\right| \approx 0.146$$

and

$$R(3,3) = \left|\sqrt{2} + \sqrt{2} - \sqrt{3} - 1\right| \approx 0.096.$$

It is easy to see that $R(n,k) \leq r_1(n,k)$ and $R(n,k) \leq r_2(n,k)$. Since we are mainly interested in the lower bounds, we shall be concentrating on $R(n,k)$. If one can show that $R(n,k) \geq 1/2^{\text{poly}(k \log n)}$, then comparing the sum of $k$ square roots of integers no larger than $n$ can be done in time polynomial in $k$ and $\log n$.

The problem of sum of square roots has recently attracted attention. First it is the main barrier to accurately classify some of the most fundamental computational problems in Euclidean space, such as the shortest path problem, the minimum spanning tree problem and the traveling salesman problem [5]. Secondly it is the simplest among the problems of the sign determination of algebraic numbers of high degree. Thirdly it has been used to show hardness of problems in other area such as approximation of 3-player Nash equilibrium [4].

### 1.1 Previous work

The zeroes of

$$f(s_1, \cdots, s_k) = \sqrt{s_1} + \sqrt{s_2} + \cdots + \sqrt{s_{\lfloor k/2 \rfloor}} - \sqrt{s_{\lfloor k/2 \rfloor + 1}} - \cdots - \sqrt{s_k}$$

form a surface in $\mathbf{R}_+^k$, where $s_i$ is nonnegative for $1 \leq i \leq k$. To bound $r_1(n,k)$ we consider an equivalent problem: how near to the surface can an integral point of absolute height no larger than $n$ get and still miss? In general finding a near-miss integral point to a surface is a very hard problem. Elkies [3] presented algorithms for these kind of problems with time complexity better than an exhaustive search. As an example, he showed how to find integral points near the curve $x^3 - y^2 = 0$. It seems hard to generalize his algorithm to the sum of square roots problem as the dimension is much higher.

The known lower bound comes from the root separation technique (for instance see [2] and [1]), which shows that

$$r_1(n,k) \geq \max\left((k\sqrt{n})^{-2^{k-1}}, (k\sqrt{n})^{-2^{\pi(n)-1}}\right)$$

where $\pi(n)$ is the number of primes no larger than $n$, and

$$R(n,k) \geq \max\left((2k\sqrt{n})^{-2^{k-1}}, (2k\sqrt{n})^{-2^{\pi(n)-1}}\right).$$

For example, it gives

$$R(165, 100) \geq \left(200\sqrt{165}\right)^{-2^{37}} \approx 10^{-468635490828}. \tag{1}$$

The lower bound is too small when $k$ and $n$ are large. However no significantly better lower bound has been reported as far as we are aware.

Qian and Wang [9] presented an upper bound for $r_1(n,k)$ based on the inequality:

$$\left|\sum_{i=0}^{k}\binom{k}{i}(-1)^i\sqrt{t+i}\right| \leq \frac{1*3*5*\cdots*(2k-3)}{2^k t^{k-\frac{1}{2}}}.$$

Note that $\binom{k}{i}$ can be as large as $\binom{k}{k/2} \geq 2^k/k$. For any fixed positive integer $k$, taking

$$n = 2^{2k}t \geq \max_{0 \leq i \leq k}\binom{k}{i}^2(t+i), \tag{2}$$

we have

$$r_1(n,k) \leq \left|\sum_{i=0}^{k}(-1)^i\sqrt{\binom{k}{i}^2(t+i)}\right|$$

$$\leq \frac{1*3*5*\cdots*(2k-3)*2^{2k^2-2k}}{(2^{2k}t)^{k-\frac{1}{2}}}$$

$$= \frac{C_k}{n^{k-\frac{1}{2}}},$$

where $C_k = 1*3*5*\cdots*(2k-3)*2^{2k^2-2k}$ is a constant depending only on $k$. By (2), we have that Qian and Wang's result only applies when $n$ is much greater than $2^{2k}$. In particular it does not give a meaningful bound when $k = 100$ and $n \leq 2^{200} \approx 10^{60}$.

## 1.2 Our results

We present a method to numerically bound $R(n,k)$ from below based on lattice reduction. Our method is efficient for large $k$ and $n$ such as $k = 100$ and $n = 165$, where an exhaustive search is clearly infeasible. The lower bounds we obtain are much better than the root separation bound. See Table 1 that compares our lower bounds with that of the root separation technique.

**Table 1.** Comparing our lower bounds with those of the root separation technique

| $R(n,k)$ | Root Separation Technique Lower Bound | Lattice Reduction Technique Lower Bound |
|---|---|---|
| $R(15, 10)$ | $10^{-60}$ | $10^{-20}$ |
| $R(33, 20)$ | $10^{-2418}$ | $10^{-50}$ |
| $R(47, 30)$ | $10^{-42832}$ | $10^{-80}$ |
| $R(66, 40)$ | $10^{-368688}$ | $10^{-120}$ |
| $R(82, 50)$ | $10^{-6201084}$ | $10^{-155}$ |
| $R(97, 60)$ | $10^{-51549123}$ | $10^{-195}$ |
| $R(113, 70)$ | $10^{-1703312763}$ | $10^{-240}$ |
| $R(131, 80)$ | $10^{-7006714363}$ | $10^{-290}$ |
| $R(146, 90)$ | $10^{-28668468036}$ | $10^{-335}$ |
| $R(165, 100)$ | $10^{-468635490828}$ | $10^{-390}$ |

Define $[x] = \lfloor x + 1/2 \rfloor$ and $\{x\} = x - [x]$. We call an integer $b$ square-free if there does not exist an integer $a > 1$ such that $a^2 | b$. We denote the $i$-th square free integer starting from 2 by $\sigma(i)$. It is known that the square roots of distinct square-free integers are linearly independent over $\mathbf{Q}$ and $\sigma(i)$ satisfies (see [8])

$$\sigma(i) = \pi^2 i/6 + O(\sqrt{i}). \qquad (3)$$

Let $s_1, s_2, \cdots, s_k$ be the distinct square-free integers no smaller than 2. Let $N$ be a positive integer. Our method is based on studying the integral lattice generated by the following $k+1$ vectors in $\mathbf{R}^{k+1}$,

$$\begin{aligned}
\mathbf{v}_0 &= (N, 0, 0, 0, \cdots, 0) \\
\mathbf{v}_1 &= ([N\sqrt{s_1}], 1, 0, 0, \cdots, 0) \\
\mathbf{v}_2 &= ([N\sqrt{s_2}], 0, 1, 0, \cdots, 0, ) \\
\mathbf{v}_3 &= ([N\sqrt{s_3}], 0, 0, 1, \cdots, 0, ) \\
&\vdots \\
\mathbf{v}_k &= ([N\sqrt{s_k}], 0, 0, 0, \cdots, 1).
\end{aligned}$$

We denote the lattice by $L_{s_1, s_2, \cdots, s_k}(N)$. If $s_1 = 2, s_2 = 3, \cdots, s_k = \sigma(k)$ are the consecutive square free integers, we will simply use $L(k, N)$ to denote the lattice.

In this paper, we are mainly concerned with $R(\sigma(k), k)$, since a good lower bound on $R(\sigma(k), k)$ can imply a good lower bound on $R(n, k)$ whenever $n = k^{O(1)}$.

**Lemma 1.** *If $R(\sigma(k), k) \geq 1/2^{\mathrm{poly}(k)}$, then $R(n, k) \geq 1/2^{\mathrm{poly}(nk)}$.*

*Proof.* If $n < \sigma(k)$, then $R(n, k) \geq R(\sigma(k), k)$. If $n > \sigma(k)$, by (3), there exists $k' = 6n/\pi^2 + O(\sqrt{n})$ such that $\sigma(k') \leq n < \sigma(k' + 1)$, then

$$R(n, k) \geq R(\sigma(k'), k') = 1/2^{\mathrm{poly}(nk)}.$$

The following theorem relates the shortest vector of $L(k, N)$ to a lower bound of $R(\sigma(k), k)$.

**Theorem 1.** *If there is a positive integer $N$ such that the shortest nonzero vector in $L(k, N)$ has length greater than $\sqrt{(1 + k\sqrt{\sigma(k)}/2)^2 + k^2 \sigma(k)}$, then*

$$R(\sigma(k), k) \geq 1/N.$$

We can also obtain constructive upper bounds from the following theorem.

**Theorem 2.** *Let $(s, a_1, a_2, \cdots, a_k)$ be a vector in $L(k, N)$. Then there exists an integer $b$ such that*

$$\left| \sum_{i=1}^{k} a_i \sqrt{\sigma(i)} - b \right| \leq \left( |s| + \sum_{i=1}^{k} |a_i|/2 \right) \frac{1}{N}.$$

We set $N$ to be large and use a lattice reduction algorithm to find a short vector $(s, a_1, a_2, \cdots, a_k)$ in the lattice $L(k, N)$. It gives us a constructive upper bound. For example, we have found that integers $a_1, a_2, \cdots, a_{100}$ and $t$ such that $\max_{1 \leq i \leq 100} a_i^2 \sigma(i) = 19796$ and

$$\left| \sum_{i=1}^{100} a_i \sqrt{\sigma(i)} - t \right| \approx 10^{-115},$$

which implies a constructive upper bound $R(19796, 100) \leq 10^{-115}$.

### 1.3 Organization

In Section 2, we review some relevant facts about lattice and present our algorithm to find a lower bound for $R(\sigma(k), k)$. In Section 3, we prove a rigorous exponential time upper bound $\exp(O(k))$ for the algorithm and present some numerical data. Based on the data we formulate a conjecture which implies that our algorithm runs in time $O(\mathrm{poly}(k))$. In Section 4, we prove Theorem 1. In Section 5, we prove Theorem 2 and another theorem on a provable upper bound for some $R(n, k)$ where $n$ is much smaller than $2^{2k}$. Throughout this paper, we use lattice functions in Victor Shoup's NTL package to produce numerical data. The block size of the BKZ reduction is set to be 10.

## 2 Lattices and Our Algorithm

In the $m$-dimensional Euclidean space $\mathbf{R}^m$, a (full rank) integral lattice is the set
$$\left\{\sum_{i=1}^m x_i \mathbf{b}_i | x_i \in \mathbf{Z}\right\},$$
where $\mathbf{b}_1, \mathbf{b}_2, \cdots \mathbf{b}_m$ are linearly independent vectors over $\mathbf{R}$ and $\mathbf{b}_i \in \mathbf{Z}^m$ for all $1 \leq i \leq m$. The determinant of a lattice is defined to be the absolute value of the determinant of the matrix $(b_{ij})$ where $b_{ij}$ is the $j$-th coordinate of $\mathbf{b}_i$. Assume that a lattice has determinant $D$ and the shortest nonzero vector has length $\lambda$. Minkowski's first theorem (see page 12 in [7]) asserts that $\lambda \leq \sqrt{m} D^{1/m}$.

Finding the shortest nonzero vector in a lattice is a well studied problem. The Block-Korkine-Zolotarev (BKZ) lattice reduction algorithm, which is based on the famous LLL lattice reduction algorithm, can find a nonzero vector whose length is at most $2^{O(m(\ln \ln m)^2 / \ln m)} \lambda$ in polynomial time [10]. Although the algorithm usually performs better than the worst case approximation ratio, it is not believed that a polynomial time algorithm can find nonzero vectors of length $2^{o(\sqrt{\log m})} \lambda$ for general lattices [6]. See [7] for a survey on computational lattice problems.

To use Theorem 1, we need a good lower bound on the length of the shortest nonzero vector in $L(k, N)$. We first apply the BKZ reduction algorithm on $L(k, N)$ to obtain a reduced base. We then apply the Gram-Schmidt orthogonalization on the reduced base. Let $\lambda^*(k, N)$ denote the length of the shortest Gram-Schmidt vector (we will omit $k$ and $N$ if they are clear from the context). Then $\lambda^*(k, N)$ is a lower bound for the length of the shortest nonzero vector in $L(k, N)$. The main process of our method can be illustrated as follows:

$$L(k, N) \overset{BKZ}{\Longrightarrow} (\mathbf{v}'_0, \mathbf{v}'_1, \cdots, \mathbf{v}'_k) \overset{Gram - Schmidt}{\Longrightarrow} (\mathbf{v}^*_0, \mathbf{v}^*_1, \cdots, \mathbf{v}^*_k).$$

Note that one should not apply the Gram-Schmidt orthogonalization directly on $L(k, N)$. Otherwise the shortest Gram-Schmidt vector will always have length 1. The algorithm is described as follows.

---

**Algorithm 1:**
Input: $k$, step;

1. $N = 1$;
2. $\lambda^* = 0$;
3. while $\lambda^* \leq \sqrt{(1 + k\sqrt{\sigma(k)}/2)^2 + k^2 \sigma(k)}$ do
4.    $N = N * step$;
5.    Apply the BKZ lattice reduction algorithm on $L(k, N)$;
6.    Apply the Gram-Schmidt orthogonalization on the reduced base;
7.    Let $\lambda^*$ be the length of the shortest vector in the Gram-Schmidt base.
8. endwhile
9. Output $1/N$ as the lower bound for $R(\sigma(k), k)$.

# 3 Time Complexity Analysis and Numerical Data

**Theorem 3.** *Algorithm 1 runs in time at most* $\exp(O(k))$.

*Proof.* Denote the length of the shortest vector in $L(k,N)$ by $\lambda$. Let $l$ be the length of the shortest vector in the reduced base. From the proof of Lemma 2.8 in [7], we derive that
$$\lambda \leq l \leq 2^{k+1}\lambda^*.$$
We shall prove that if $N \geq 2^{3k2^k}$, then $\lambda \geq 2^{2k}$, which implies that for $k > 7$
$$\lambda^* \geq \frac{\lambda}{2^{k+1}} \geq 2^{k-1} > 3k^{1.5}.$$
On the other hand, we know from formulae (3) that for $k$ big enough, $\sigma(k) < 2k$. Hence
$$\sqrt{(1+k\sqrt{\sigma(k)}/2)^2 + k^2\sigma(k)} < \sqrt{(1+k\sqrt{2k}/2)^2 + 2k^3} < \sqrt{2k^3 + 2k^3} < 3k^{1.5}.$$
This shows that the algorithm will terminate before $N$ exceeds $2^{3k2^k}$. The time complexity is thus at most $O\left(k2^k \text{poly}\left(k \log 2^{3k2^k}\right)\right)$, which is $\exp(O(k))$.

Assume that $N \geq 2^{3k2^k}$. Any nonzero vector in the lattice has form
$$\left(\sum_{i=1}^{k} a_i \left[N\sqrt{\sigma(i)}\right] - bN, a_1, a_2, \cdots, a_k\right)$$
for some integers $a_1, a_2, \cdots, a_k$ and $b$. It is enough to show that the length of the vector is greater than $2^{2k}$. If for some $a_i$, $|a_i| > 2^{2k}$, then the length of the vector is greater than $2^{2k}$. So we may assume that $|a_i| \leq 2^{2k}$ for all $1 \leq i \leq k$.

$$\left|\sum_{i=1}^{k} a_i \left[N\sqrt{\sigma(i)}\right] - bN\right| = \left|\sum_{i=1}^{k} a_i \left(N\sqrt{\sigma(i)} - \left\{N\sqrt{\sigma(i)}\right\}\right) - bN\right|$$
$$\geq \left|\sum_{i=1}^{k} a_i N\sqrt{\sigma(i)} - bN\right| - \left|\sum_{i=1}^{k} a_i \left\{N\sqrt{\sigma(i)}\right\}\right|$$
$$\geq N\left|\sum_{i=1}^{k} a_i\sqrt{\sigma(i)} - b\right| - \sum_{i=1}^{k} \frac{|a_i|}{2}$$
$$\geq N\left|\sum_{i=1}^{k} a_i\sqrt{\sigma(i)} - b\right| - \frac{k2^{2k}}{2}$$

By the root separation bound,
$$N\left|\sum_{i=1}^{k} a_i\sqrt{\sigma(i)} - b\right| - \frac{k2^{2k}}{2} \geq N\left(2k\sqrt{2^{4k}\sigma(k)}\right)^{-2^{k-1}} - \frac{k2^{2k}}{2}$$
which is greater than $2^{2k}$ as $N \geq 2^{3k2^k}$.

The above theorem gives us an exponential upper bound of the time complexity. However from numerical experiments, we can see that the algorithm terminates quickly and enables us to find a lower bound of $R(\sigma(k), k)$ much better than the root separation bound.

We list in Table 2 the values of $l^2$ and $(\lambda^*)^2$ for $L(100, N)$ where $N$ starts from $10^{50}$ and keeps increasing by a factor of $10^5$. From Table 2, we learn that the square length of the shortest nonzero vector in the lattice $L(100, 10^{390})$ is greater than 3102794. Since $\left(1 + 100\sqrt{165}/2\right)^2 + 100^2 * 165 = 2063785.52..$, we obtain that $R(165, 100) \geq 10^{-390}$. Similarly we can get the other data on the right-hand side in Table 1.

Table 2 and Table 3 illustrate that the ratio between $\lambda^*$ and $N^{\frac{1}{k+1}}$ remains about the same when $N$ increases. Note that $\lambda^*(k, N)$ is the lower bound of the length of the shortest nonzero vector in $L(k, N)$. Based on this observation, we formulate the following conjecture on the length of the shortest nonzero vector in $L(k, N)$:

*Conjecture 1.* The shortest nonzero vector in the lattice $L(k, N)$ has length greater than $N^{\frac{1}{k+1}}/k$.

**Corollary 1.** *If Conjecture 1 is true then*

1. $R(\sigma(k), k) \geq 1/(2\sigma(k)k^3)^k$,
2. *Algorithm 1 runs in time $O(\text{poly}(k))$.*

*Proof.* Set
$$N = \left\lceil \left( k \left( \left(1 + k\sqrt{\sigma(k)}/2\right)^2 + k^2\sigma(k) \right) \right)^k \right\rceil.$$

The first item follows from Theorem 1 and Conjecture 1. The time complexity of Algorithm 1 is $O(k \cdot \text{poly}(k \log N))$, which is at most $O(\text{poly}(k))$. Thus the second item holds.

It is interesting to contrast $L(k, N)$ with a similar lattice generated by

$$\begin{aligned}
&(N, 0, 0, 0, \cdots, 0) \\
&(N[\sqrt{s_1}], 1, 0, 0, \cdots, 0) \\
&(N[\sqrt{s_2}], 0, 1, 0, \cdots, 0,) \\
&(N[\sqrt{s_3}], 0, 0, 1, \cdots, 0,) \\
&\vdots \\
&(N[\sqrt{s_k}], 0, 0, 0, \cdots, 1).
\end{aligned}$$

It is easy to see that the shortest vector in the lattice always has length 1 no matter how large $N$ is.

**Table 2.** The data for $L(100, N)$ ($\sigma(100) = 165$)

| $\log_{10} N$ | $l^2$ | $(\lambda^*)^2$ | $\lambda^*/N^{\frac{1}{k+1}}$ | $\log_{10} N$ | $l^2$ | $(\lambda^*)^2$ | $\lambda^*/N^{\frac{1}{k+1}}$ |
|---:|---:|---:|---:|---:|---:|---:|---:|
| 50 | 189 | 0.920102 | 0.31 | 55 | 187 | 1 | 0.29 |
| 60 | 267 | 0.848578 | 0.23 | 65 | 318 | 1.075123 | 0.24 |
| 70 | 379 | 1.251328 | 0.23 | 75 | 531 | 1.496321 | 0.22 |
| 80 | 711 | 1.905386 | 0.22 | 85 | 824 | 2.645646 | 0.23 |
| 90 | 1039 | 3.06967 | 0.23 | 95 | 1466 | 3.489959 | 0.21 |
| 100 | 1959 | 5.242274 | 0.23 | 105 | 2339 | 6.414824 | 0.23 |
| 110 | 2726 | 7.750273 | 0.23 | 115 | 3639 | 10.30219 | 0.23 |
| 120 | 4370 | 11.97402 | 0.22 | 125 | 5512 | 14.85322 | 0.22 |
| 130 | 6936 | 17.7999 | 0.22 | 135 | 9345 | 26.00225 | 0.23 |
| 140 | 10479 | 32.10648 | 0.23 | 145 | 11789 | 38.72952 | 0.23 |
| 150 | 18949 | 45.87119 | 0.22 | 155 | 20579 | 71.66641 | 0.25 |
| 160 | 23457 | 81.06815 | 0.23 | 165 | 33572 | 108.6672 | 0.24 |
| 170 | 40148 | 122.6395 | 0.23 | 175 | 52839 | 157.6955 | 0.23 |
| 180 | 72509 | 185.265 | 0.22 | 185 | 81229 | 271.003 | 0.24 |
| 190 | 96242 | 279.2017 | 0.22 | 195 | 116002 | 416.3477 | 0.24 |
| 200 | 165201 | 421.7492 | 0.21 | 205 | 182891 | 662.0805 | 0.24 |
| 210 | 267509 | 786.044 | 0.23 | 215 | 307450 | 1103.5 | 0.25 |
| 220 | 411290 | 1241.363 | 0.23 | 225 | 530717 | 1566.732 | 0.23 |
| 230 | 484931 | 1948.158 | 0.23 | 235 | 761232 | 2508.116 | 0.24 |
| 240 | 1010500 | 2877.924 | 0.23 | 245 | 1273090 | 3674.155 | 0.23 |
| 250 | 1628420 | 4691.54 | 0.23 | 255 | 1699623 | 5815.91 | 0.23 |
| 260 | 2345069 | 8097.233 | 0.24 | 265 | 2735544 | 10196.72 | 0.24 |
| 270 | 3830216 | 12159.34 | 0.23 | 275 | 4483731 | 14841.72 | 0.23 |
| 280 | 6448489 | 20310.26 | 0.24 | 285 | 7963507 | 21892.18 | 0.22 |
| 290 | 9576142 | 29173.78 | 0.23 | 295 | 11065095 | 37113.27 | 0.23 |
| 300 | 14625831 | 47887.96 | 0.23 | 305 | 20017801 | 54521.73 | 0.22 |
| 310 | 22107891 | 70218.28 | 0.23 | 315 | 30329692 | 92013.94 | 0.23 |
| 320 | 42326718 | 133759.6 | 0.25 | 325 | 50110184 | 160807.8 | 0.24 |
| 330 | 58226453 | 163842 | 0.22 | 335 | 73620063 | 209813 | 0.22 |
| 340 | 101230523 | 289148.6 | 0.23 | 345 | 116341856 | 367148.5 | 0.23 |
| 350 | 134263638 | 446526.2 | 0.23 | 355 | 176973638 | 606588.9 | 0.24 |
| 360 | 195623258 | 577848.5 | 0.21 | 365 | 295497369 | 953276.7 | 0.24 |
| 370 | 254486097 | 1108836 | 0.23 | 375 | 365813532 | 1383326 | 0.23 |
| 380 | 620569774 | 1500616 | 0.21 | 385 | 857210733 | 1906762 | 0.21 |
| 390 | 936892309 | 3102794 | 0.24 | 395 | 1214701229 | 3716443 | 0.24 |
| 400 | 1512222196 | 4440063 | 0.23 | 405 | 1815150428 | 4762491 | 0.21 |
| 410 | 2479097817 | 6683891 | 0.23 | 415 | 2825781352 | 8365474 | 0.23 |
| 420 | 3315764769 | 11230408 | 0.23 | 425 | 4759873617 | 13620881 | 0.23 |
| 430 | 5567641140 | 18494986 | 0.24 | 435 | 7323262414 | 20738831 | 0.22 |
| 440 | 9416436554 | 29562467 | 0.24 | 445 | 10529606845 | 31706291 | 0.22 |
| 450 | 13988924828 | 44253095 | 0.23 | 455 | 20051717365 | 58023065 | 0.24 |
| 460 | 24926540282 | 72675194 | 0.24 | 465 | 26802599860 | 84410497 | 0.23 |
| 470 | 35908570410 | 88432810 | 0.21 | 475 | 42772055636 | 1.46E+08 | 0.24 |
| 480 | 55777952874 | 1.55E+08 | 0.22 | 485 | 68743792860 | 2.23E+08 | 0.24 |
| 490 | 89063811044 | 2.9E+08 | 0.24 | 495 | 1.18668E+11 | 3.22E+08 | 0.23 |
| 500 | 1.37029E+11 | 4.57E+08 | 0.24 | 505 | 1.79821E+11 | 5.16E+08 | 0.23 |
| 510 | 2.02652E+11 | 7.26E+08 | 0.24 | 515 | 2.5021E+11 | 7.87E+08 | 0.22 |
| 520 | 4.04579E+11 | 1.12E+09 | 0.24 | 525 | 4.64658E+11 | 1.4E+09 | 0.24 |
| 530 | 5.00824E+11 | 1.6E+09 | 0.23 | 535 | 6.7777E+11 | 2.51E+09 | 0.25 |
| 540 | 9.40142E+11 | 2.4E+09 | 0.22 | 545 | 1.03272E+12 | 3.12E+09 | 0.22 |
| 550 | 1.23394E+12 | 4.13E+09 | 0.23 | 555 | 1.7404E+12 | 5.28E+09 | 0.23 |
| 560 | 2.04264E+12 | 6.78E+09 | 0.23 | 565 | 2.53027E+12 | 7.92E+09 | 0.23 |
| 570 | 3.14243E+12 | 1.09E+10 | 0.24 | 575 | 4.00544E+12 | 1.29E+10 | 0.23 |
| 580 | 5.82342E+12 | 1.37E+10 | 0.21 | 585 | 6.91689E+12 | 2.09E+10 | 0.23 |
| 590 | 8.50935E+12 | 2.13E+10 | 0.21 | 595 | 1.11703E+13 | 2.95E+10 | 0.22 |

**Table 3.** The data for $\lambda^*(k,N)/N^{\frac{1}{k+1}}$

| $k$ \ $\log_{10} N$ | 10 | 20 | 30 | 40 | 50 | 60 | 70 | 80 | 90 | 100 |
|---|---|---|---|---|---|---|---|---|---|---|
| 50  | 0.83 | 0.74 | 0.62 | 0.56 | 0.48 | 0.42 | 0.36 | 0.29 | 0.27 | 0.31 |
| 60  | 0.91 | 0.78 | 0.64 | 0.57 | 0.47 | 0.41 | 0.34 | 0.32 | 0.27 | 0.23 |
| 70  | 0.90 | 0.78 | 0.65 | 0.55 | 0.47 | 0.39 | 0.35 | 0.31 | 0.29 | 0.23 |
| 80  | 0.91 | 0.77 | 0.66 | 0.58 | 0.50 | 0.42 | 0.36 | 0.33 | 0.28 | 0.22 |
| 90  | 0.90 | 0.69 | 0.65 | 0.58 | 0.49 | 0.41 | 0.37 | 0.30 | 0.24 | 0.23 |
| 100 | 0.90 | 0.76 | 0.65 | 0.57 | 0.49 | 0.43 | 0.35 | 0.30 | 0.28 | 0.23 |
| 110 | 0.94 | 0.73 | 0.64 | 0.56 | 0.46 | 0.44 | 0.35 | 0.32 | 0.27 | 0.23 |
| 120 | 0.75 | 0.73 | 0.62 | 0.59 | 0.47 | 0.41 | 0.34 | 0.30 | 0.27 | 0.22 |
| 130 | 0.91 | 0.80 | 0.63 | 0.56 | 0.49 | 0.42 | 0.36 | 0.29 | 0.27 | 0.22 |
| 140 | 0.89 | 0.78 | 0.67 | 0.56 | 0.51 | 0.39 | 0.35 | 0.31 | 0.29 | 0.23 |
| 150 | 0.89 | 0.77 | 0.66 | 0.53 | 0.48 | 0.43 | 0.35 | 0.32 | 0.27 | 0.22 |
| 160 | 0.92 | 0.79 | 0.64 | 0.59 | 0.50 | 0.39 | 0.36 | 0.32 | 0.25 | 0.23 |
| 170 | 0.89 | 0.75 | 0.64 | 0.51 | 0.47 | 0.41 | 0.36 | 0.31 | 0.28 | 0.23 |
| 180 | 0.92 | 0.74 | 0.65 | 0.58 | 0.48 | 0.43 | 0.36 | 0.30 | 0.25 | 0.22 |
| 190 | 0.93 | 0.84 | 0.65 | 0.56 | 0.45 | 0.38 | 0.39 | 0.29 | 0.26 | 0.22 |
| 200 | 0.90 | 0.78 | 0.66 | 0.52 | 0.47 | 0.42 | 0.34 | 0.32 | 0.28 | 0.21 |
| 210 | 0.83 | 0.80 | 0.64 | 0.56 | 0.48 | 0.42 | 0.40 | 0.33 | 0.25 | 0.23 |
| 220 | 0.79 | 0.77 | 0.67 | 0.57 | 0.51 | 0.44 | 0.35 | 0.30 | 0.28 | 0.23 |
| 230 | 0.94 | 0.73 | 0.62 | 0.58 | 0.44 | 0.43 | 0.33 | 0.30 | 0.27 | 0.23 |
| 240 | 0.91 | 0.79 | 0.67 | 0.58 | 0.48 | 0.42 | 0.36 | 0.31 | 0.25 | 0.23 |
| 250 | 0.91 | 0.80 | 0.63 | 0.53 | 0.47 | 0.41 | 0.36 | 0.31 | 0.26 | 0.23 |
| 260 | 0.93 | 0.77 | 0.63 | 0.55 | 0.47 | 0.43 | 0.36 | 0.33 | 0.28 | 0.24 |
| 270 | 0.91 | 0.74 | 0.66 | 0.54 | 0.50 | 0.43 | 0.39 | 0.30 | 0.27 | 0.23 |
| 280 | 0.85 | 0.81 | 0.63 | 0.55 | 0.47 | 0.42 | 0.36 | 0.31 | 0.26 | 0.24 |
| 290 | 0.92 | 0.79 | 0.64 | 0.55 | 0.49 | 0.40 | 0.37 | 0.32 | 0.26 | 0.23 |
| 300 | 0.94 | 0.78 | 0.68 | 0.57 | 0.47 | 0.42 | 0.34 | 0.32 | 0.25 | 0.23 |

# 4 The Proof of Theorem 1

To prove Theorem 1, we need the following lemma.

**Lemma 2.** *Let $s_1, s_2, \cdots, s_k$ be $k$ distinct positive square free integers. Let $\lambda$ be the length of the shortest nonzero vector in $L_{s_1,s_2,\cdots,s_k}(N)$. For any integers $a_1, a_2, \cdots, a_k, b$, if $(b, a_1, a_2, \cdots, a_k) \neq (0, 0, 0, \cdots, 0)$, and $\lambda^2 \geq \left(1 + \frac{\sum_{i=1}^{k} |a_i|}{2}\right)^2 + \sum_{i=1}^{k} a_i^2$, then*

$$\left|\sum_{i=1}^{k} a_i \sqrt{s_i} - b\right| \geq \frac{1}{N}.$$

*Proof.* The vector

$$\left(\sum_{i=1}^{k} a_i[N\sqrt{s_i}] - bN, a_1, a_2, \cdots, a_k\right)$$

is nonzero and in the lattice, hence its length

$$\sqrt{\sum_{i=1}^{k} a_i^2 + \left(\sum_{i=1}^{k} a_i[N\sqrt{s_i}] - bN\right)^2}$$

is no smaller than $\lambda$. We have

$$\sum_{i=1}^{k} a_i^2 + \left(\sum_{i=1}^{k} a_i[N\sqrt{s_i}] - bN\right)^2 \geq \lambda^2 \geq \left(1 + \frac{\sum_{i=1}^{k} |a_i|}{2}\right)^2 + \sum_{i=1}^{k} a_i^2.$$

It implies that

$$\left|\sum_{i=1}^{k} a_i[N\sqrt{s_i}] - bN\right| \geq 1 + \frac{\sum_{i=1}^{k} |a_i|}{2}.$$

The left hand side is

$$\left|\sum_{i=1}^{k} a_i[N\sqrt{s_i}] - bN\right| = \left|\sum_{i=1}^{k} a_i(N\sqrt{s_i} - \{N\sqrt{s_i}\}) - bN\right|$$

$$\leq \left|\sum_{i=1}^{k} a_i N\sqrt{s_i} - bN\right| + \left|\sum_{i=1}^{k} a_i\{N\sqrt{s_i}\}\right|$$

$$\leq \left|\sum_{i=1}^{k} a_i N\sqrt{s_i} - bN\right| + \sum_{i=1}^{k} |a_i\{N\sqrt{s_i}\}|$$

$$\leq \left|\sum_{i=1}^{k} a_i N\sqrt{s_i} - bN\right| + \frac{\sum_{i=1}^{k} |a_i|}{2}$$

So we have
$$\left|\sum_{i=1}^{k} a_i N\sqrt{s_i} - bN\right| \geq 1,$$
therefore $\left|\sum_{i=1}^{k} a_i\sqrt{s_i} - b\right| \geq 1/N$.

Now we are ready to prove Theorem 1.

*Proof.* Let $n_i, 1 \leq i \leq k$ be positive integers $\leq \sigma(k)$, $m$ be an integer and $e_i \in \{1, 0, -1\}$ for all $1 \leq i \leq k$. We can write $\sum_{i=1}^{k} e_i\sqrt{n_i} - m$ as $\sum_{i=1}^{k} a_i\sqrt{\sigma(i)} - b$ where $a_1, a_2, \cdots, a_k, b$ are integers. We have that
$$\sum_{i=1}^{k} |a_i| \leq \sum_{i=1}^{k} |a_i|\sqrt{\sigma(i)} \leq \sum_{i=1}^{k} \sqrt{n_i} \leq k\sqrt{\sigma(k)}$$
and
$$\sum_{i=1}^{k} a_i^2 \leq \left(\sum_{i=1}^{k} |a_i|\right)^2 \leq k^2\sigma(k).$$
Assume that $(a_1, a_2, \cdots, a_k, b) \neq (0, 0, \cdots, 0, 0)$. Since the shortest nonzero vector in the lattice $L(k, N)$ has length at least
$$\left(1 + k\sqrt{\sigma(k)}/2\right)^2 + k^2\sigma(k) \geq \left(1 + \frac{\sum_{i=1}^{k} |a_i|}{2}\right)^2 + \sum_{i=1}^{k} a_i^2,$$
we conclude from Lemma 2 that,
$$\left|\sum_{i=1}^{k} e_i\sqrt{n_i} - m\right| = \left|\sum_{i=1}^{k} a_i\sqrt{\sigma(i)} - b\right| \geq \frac{1}{N}.$$

## 5  Upper Bound

Now we can prove Theorem 2.

*Proof.* Since $(s, a_1, a_2, \cdots, a_k)$ is a vector in the lattice $L(k, N)$, there exists an integer $b$ such that
$$\left|\sum_{i=1}^{k} a_i \left[N\sqrt{\sigma(i)}\right] - bN\right| = |s|.$$
Then
$$\left|\sum_{i=1}^{k} a_i N\sqrt{\sigma(i)} - bN\right| = \left|\sum_{i=1}^{k} a_i \left(\left[N\sqrt{\sigma(i)}\right] + \left\{N\sqrt{\sigma(i)}\right\}\right) - bN\right|$$
$$\leq \left|\sum_{i=1}^{k} a_i \left[N\sqrt{\sigma(i)}\right] - bN\right| + \left|\sum_{i=1}^{k} a_i \left\{N\sqrt{\sigma(i)}\right\}\right|$$
$$\leq |s| + \sum_{i=1}^{k} |a_i|/2.$$

Hence the theorem follows.

We may apply the BKZ reduction algorithm on the lattice and obtain a nonzero vector $(s, a_1, a_2, \cdots, a_k)$ of length at most $2^{O(k(\ln \ln k)^2/\ln k)} N^{\frac{1}{k+1}}$. The data is listed in Table 4. More generally, we have

**Theorem 4.** *Let $s_1, s_2, \cdots, s_k$ be $k$ distinct square free integers no smaller than 2. For any integer $N$, we can find integers $a_1, a_2, \cdots, a_k$ and $b$ in polynomial time satisfying that $|a_i| \leq 2^{O(k(\ln \ln k)^2/\ln k)} N^{\frac{1}{k+1}}$ for all $1 \leq i \leq k$ and*

$$\left| \sum_{i=1}^{k} a_i \sqrt{s_i} - b \right| \leq 2^{O(k(\ln \ln k)^2/\ln k)} N^{\frac{-k}{k+1}}.$$

*Proof.* The determinant of the lattice $L(k, N)$ is

$$\begin{vmatrix} N & 0 & 0 & 0 & \cdots & 0 \\ [N\sqrt{s_1}] & 1 & 0 & 0 & \cdots & 0 \\ [N\sqrt{s_2}] & 0 & 1 & 0 & \cdots & 0 \\ [N\sqrt{s_3}] & 0 & 0 & 1 & \cdots & 0 \\ & & \vdots & & & \\ [N\sqrt{s_k}] & 0 & 0 & 0 & \cdots & 1 \end{vmatrix} = N$$

By Minkowski's first theorem, there is a vector of length $\sqrt{k+1} N^{\frac{1}{k+1}}$ or shorter in the lattice. If we apply the BKZ reduction algorithm on the lattice, we obtain a nonzero vector $(s, a_1, a_2, \cdots, a_k)$ of length at most $2^{O(k(\ln \ln k)^2/\ln k)} N^{\frac{1}{k+1}}$. Thus $|a_i| \leq 2^{O(k(\ln \ln k)^2/\ln k)} N^{\frac{1}{k+1}}$ for all $1 \leq i \leq k$ and $|s| \leq 2^{O(k(\ln \ln k)^2/\ln k)} N^{\frac{1}{k+1}}$. We have

$$\left| \sum_{i=1}^{k} a_i \sqrt{s_i} - b \right| \leq \left( |s| + \sum_{i=1}^{k} |a_i|/2 \right) N^{-1} = 2^{O(k(\ln \ln k)^2/\ln k)} N^{\frac{-k}{k+1}}.$$

## 6 Concluding Remarks

In this paper we present a numerical method that finds a much better lower bound for $R(n, k)$ than the previously known methods do. The main open problem is to prove Conjecture 1, which implies that our method runs in polynomial time.

### References

1. C. Burnikel, R. Fleischer, K. Mehlhorn, and S. Schirra. A strong and easily computable separation bound for arithmetic expressions involving radicals. *Algorithmica*, 27(1):87–99, 2000.

**Table 4.** The data for the shortest vector in the BKZ reduced base: $n = \max_i a_i^2 \sigma(i)$

| $\log_{10} N$ | $n$ | $s$ | $\log_{10} N$ | $n$ | $s$ |
|---:|---:|---:|---:|---:|---:|
| 50 | 927 | 1 | 55 | 2275 | -1 |
| 60 | 1616 | -2 | 65 | 1680 | 0 |
| 70 | 3430 | 2 | 75 | 4075 | -1 |
| 80 | 7595 | -7 | 85 | 6016 | -1 |
| 90 | 7693 | -1 | 95 | 10406 | -1 |
| 100 | 21744 | -1 | 105 | 8107 | 0 |
| 110 | 13310 | 7 | 115 | 19796 | -4 |
| 120 | 25650 | 7 | 125 | 39680 | 4 |
| 130 | 43639 | 1 | 135 | 64375 | 14 |
| 140 | 88556 | 6 | 145 | 104247 | -9 |
| 150 | 144900 | 4 | 155 | 128625 | 7 |
| 160 | 427228 | -15 | 165 | 287550 | -5 |
| 170 | 236192 | 12 | 175 | 554429 | 4 |
| 180 | 543600 | 13 | 185 | 1422596 | 18 |
| 190 | 613965 | -7 | 195 | 739640 | 55 |
| 200 | 1056440 | 31 | 205 | 1609650 | 1 |
| 210 | 1417939 | 42 | 215 | 2003760 | -52 |
| 220 | 2321776 | 136 | 225 | 4790753 | 65 |
| 230 | 3819232 | 41 | 235 | 5270427 | 18 |
| 240 | 6951744 | 8 | 245 | 10285412 | 80 |
| 250 | 13564142 | -205 | 255 | 14948504 | 232 |
| 260 | 12803364 | -125 | 265 | 23483741 | -187 |
| 270 | 19553816 | 122 | 275 | 27059175 | 61 |
| 280 | 43818180 | 480 | 285 | 49650120 | 210 |
| 290 | 90805809 | 474 | 295 | 66427398 | -256 |
| 300 | 94303440 | 42 | 305 | 282685488 | -71 |
| 310 | 217669834 | 570 | 315 | 208586875 | -203 |
| 320 | 282304440 | 257 | 325 | 340940097 | 288 |
| 330 | 684191361 | -451 | 335 | 572956706 | 134 |
| 340 | 380583760 | -2671 | 345 | 522838807 | 434 |
| 350 | 2598879987 | -1205 | 355 | 1209416149 | -1467 |
| 360 | 1481771275 | 743 | 365 | 1496653524 | 1917 |
| 370 | 1564223967 | -2843 | 375 | 2345301325 | 3085 |
| 380 | 3944948022 | 135 | 385 | 4095144375 | -368 |
| 390 | 8242412940 | -2549 | 395 | 6185090625 | -1475 |
| 400 | 20837120600 | -3524 | 405 | 16962130500 | 649 |
| 410 | 35444689191 | 1235 | 415 | 16992489353 | -4827 |
| 420 | 52222173525 | 926 | 425 | 29831928284 | 5987 |
| 430 | 32518315323 | -8686 | 435 | 31432393528 | 263 |
| 440 | 97378652871 | -20156 | 445 | 58208231178 | 14650 |
| 450 | 1.1644E+11 | -4583 | 455 | 1.17189E+11 | -121 |
| 460 | 1.56626E+11 | -24968 | 465 | 2.31065E+11 | 11314 |
| 470 | 2.40379E+11 | 6759 | 475 | 1.88347E+11 | -1784 |
| 480 | 3.71914E+11 | 12199 | 485 | 5.79326E+11 | 25529 |
| 490 | 6.61439E+11 | 15283 | 495 | 7.70885E+11 | 55710 |
| 500 | 9.82574E+11 | 39390 | 505 | 1.03675E+12 | 53574 |
| 510 | 1.35333E+12 | -74474 | 515 | 2.08272E+12 | 43145 |
| 520 | 2.46576E+12 | -18596 | 525 | 3.51867E+12 | -88646 |
| 530 | 3.53844E+12 | -154254 | 535 | 7.71874E+12 | 44316 |
| 540 | 6.34504E+12 | 12527 | 545 | 5.00231E+12 | 66480 |
| 550 | 2.0875E+13 | -50539 | 555 | 1.49025E+13 | 70127 |
| 560 | 1.63683E+13 | -10313 | 565 | 4.2508E+13 | 24836 |
| 570 | 4.06111E+13 | 202903 | 575 | 2.54605E+13 | 63494 |
| 580 | 3.43017E+13 | 171277 | 585 | 7.91203E+13 | 135561 |
| 590 | 7.65752E+13 | 15225 | 595 | 8.47789E+13 | 427965 |

2. Qi Cheng. On comparing sums of square roots of small integers. In *Proc. of 31st International Symposium on Mathematical Foundations of Computer Science (MFCS)*, volume 4162 of *Lecture Notes in Computer Science*, 2006.
3. Noam D. Elkies. Rational points near curves and small nonzero $|x^3-y^2|$ via lattice reduction. In W. Bosma, editor, *Proceedings of ANTS-4*, volume 1838 of *Lecture Notes in Computer Science*, pages 33–63, 2000.
4. Kousha Etessami and Mihalis Yannakakis. On the complexity of nash equilibria and other fixed points. In *Proceedings of the 48th FOCS*, pages 113–123, 2007.
5. M. Garey, R.L. Graham, and D.S. Johnson. Some NP-complete geometric problems. In *Proc. ACM Symp. Theory Comp.*, pages 10–21, 1976.
6. Subhash Khot. Hardness of approximating the shortest vector problem in lattices. In *Proc. 45th IEEE Symp. on Foundations of Comp. Science*, 2004.
7. Daniele Micciancio and Shafi Goldwasser. *Complexity of Lattice Problems: A Cryptographic Perspective*. Kluwer Academic Publishers, 2002.
8. Francesco Pappalardi. A survey on k-power freeness. In *Proceeding of the Conference in Analytic Number Theory in Honor of Prof. Subbarao*, number 1 in Ramanujan Math. Soc. Lect. Notes Ser., pages 71–88, 2002.
9. Jianbo Qian and Cao An Wang. How much precision is needed to compare two sums of square roots of integers? *Inf. Process. Lett.*, 100(5):194–198, 2006.
10. C. P. Schnorr. A hierarchy of polynomial time lattice basis reduction algorithms. *Theoretical Computer Science*, 53(2-3):201–224, 1987.
11. Prasoon Tiwari. A problem that is easier to solve on the unit-cost algebraic RAM. *Journal of Complexity*, 8(4):393–397, 1992.